\documentclass[twocolumn,showpacs,preprintnumbers,amsmath,amssymb,mathptmx,citeautoscript,aps,prl]{revtex4-2}

\usepackage{txfonts}
\usepackage{graphicx}
\usepackage{bm}
\usepackage{color}
\usepackage{ulem}
\usepackage{cancel}
\usepackage{enumerate}

\newcommand{\jtitle}[1]{#1}

\begin{document}

\title{Enhanced Seebeck coefficient through the magnetic fluctuations in Sr$_2$Ru$_{1-x}M_x$O$_4$ ($M = $ Co, Mn)}

\author{Takayoshi Yamanaka}
\email{takayoshi.yamanaka.b5@tohoku.ac.jp}
\altaffiliation[Present address: ]{Institute for Materials Research, Tohoku University, Sendai 980‑8577, Japan}
\author{Ryuji Okazaki}
\email{okazaki@rs.tus.ac.jp}
\author{Hiroshi Yaguchi}

\affiliation{
Department of Physics, Faculty of Science and Technology, Tokyo University of Science, Noda, Chiba 278-8510, Japan
}

\date{\today}

\begin{abstract}
The layered perovskite Sr$_2$RuO$_4$ is a most intensively studied superconductor, but 
its pairing mechanism, which is often coupled intimately with magnetic fluctuations in correlated materials, is still an open question.
Here 
we present a systematic evolution of the Seebeck coefficient
in Co- and Mn-substituted Sr$_2$RuO$_4$ single crystals, 
in which ferromagnetic and antiferromagnetic glassy states respectively emerge 
in proximity to the superconducting phase of the parent compound.
We find that the Seebeck coefficient $S$ divided by temperature $T$, $S/T$, shows a maximum near 
characteristic temperatures seen in the irreversible magnetization $M_{\rm ir}$
in both of the Co- and Mn-substituted crystals,
demonstrating both of the ferromagnetic and antiferromagnetic fluctuations to enhance the Seebeck coefficient.
Interestingly, $S/T$ increases with lowering temperature in the parent compound,
reminiscent of non-Fermi-liquid behavior,
indicating an essential role of coexisting ferromagnetic and antiferromagnetic fluctuations for the itinerant electrons in Sr$_2$RuO$_4$.
\end{abstract}

\maketitle

\section{I. Introduction}

Unconventional superconductivity in the layered perovskite Sr$_2$RuO$_4$
has long attracted interest \cite{Maeno1994,Rice1995,Mackenzie2003,Kallin2012,Mackenzie2017,Kivelson2020}.
In particular, recent progress of nuclear magnetic resonance (NMR) experiments
has posed strong constraints on the spin sector of the superconducting order parameter \cite{Pustogow2019,Ishida2020,Chronister2021},
offering
a clue for the consistent understanding of several experimental results \cite{Kittaka2009,Yonezawa2013,Hicks2014}
that
were difficult to reconcile with earlier NMR results \cite{Ishida1998,Murakawa2007,Ishida2008,Ishida2015},
as well as stimulating further symmetry-based experiments to elucidate the order parameter \cite{Benhabib2021,Ghosh2021,Grinenko2021}.
As a result, an exotic two-component order parameter with broken time reversal symmetry has been proposed,
which is unique as suggested in several examples \cite{Kasahara2007,Smidman2015},
and various theoretical attempts have also been made \cite{Gingras2019,Romer2019,Suh2020,Willa2021,Romer2021,Zhang2021},
opening an avenue for exploring unconventional pairing interaction to realize such order parameters.

To address this underlying issue, it is crucially important to establish the electronic phase diagram
as a function of external parameters such as pressure and chemical substitutions,
as is widely discussed in correlated matters \cite{Das2016, Paschen2021}.
Indeed, although the superconducting state in Sr$_2$RuO$_4$ is extremely sensitive to impurity \cite{Mackenzie1998}, 
dramatic changes in the electronic state with elemental substitutions have been reported.
In the isovalent systems, for instance, 
a spin-glass state develops over a wide range of the Ca content in (Sr, Ca)$_2$RuO$_4$ \cite{Nakatsuji2000,Carlo2012},
which may originate from the degree of freedom in the RuO$_6$ octahedra.
On the other hand,
in Sr$_2$(Ru, Ti)O$_4$,
an incommensurate spin-density-wave ordering due to the Fermi-surface nesting appears with glassy behavior \cite{Minakata2001}.
In contrast to the antiferromagnetic (AFM) coupling seen in the isovalent systems, 
La$^{3+}$ substitution to Sr$^{2+}$ sites results in electron doping 
to expand the electron-like $\gamma$ Fermi surface,
leading to ferromagnetic (FM) fluctuation
owing to an enhancement of the density of states (DOS)
at the Fermi energy $N(\varepsilon_{\rm F})$ 
near the van Hove singularity (vHs) \cite{Kikugawa2004_La,Kikugawa2004,Shen2007}.
These results clearly show the complicated magnetic instabilities existing in the parent compound Sr$_2$RuO$_4$,
which involve complex structural and electronic origins.
Such instabilities are also discussed in several studies
including NMR and neutron experiments \cite{Imai1998,Sidis1999,Braden2002,Steffens2019,Jenni2021}.

Among many substituted systems,
Co- and Mn-substituted Sr$_2$RuO$_4$ serve as a
fascinating platform to investigate 
how magnetic fluctuations mediate the emergence of superconductivity,
because slight substitutions of Co and Mn drastically vary the superconducting ground state into
the FM and AFM glassy states,
respectively \cite{Ortmann2013}.
In the Co-substituted system, 
the FM cluster glass characterized by an exponential relaxation of the remanent magnetization appears at low temperatures.
Also, the electronic specific heat $\gamma_{\rm e}$ increases with increasing Co contents,
indicating an increased $N(\varepsilon_{\rm F})$ similar to La-substitution \cite{Kikugawa2004_La}.
It is noteworthy that 
the increase of $N(\varepsilon_{\rm F})$ in a $1.5\%$ Co-substituted sample 
estimated from $\gamma_{\rm e}$ is comparable to 
that in a $5\%$ La-substituted sample,
possibly implying an effective electron doping effect by the Co substitution \cite{Ortmann2013}.
Similarly, the AFM transition temperature in Mn-substituted samples is 
much higher than that in Ti-substituted ones,
demonstrating enigmatic roles of the Co and Mn substitutions 
to strengthen the magnetic couplings in Sr$_2$RuO$_4$.

The aim of this study is to examine such magnetic fluctuations in 
Sr$_2$Ru$_{1-x}M_x$O$_4$ ($M = $ Co, Mn)
by means of Seebeck coefficient measurement,
known as 
a powerful tool for fluctuations
as it is a measure of the entropy per charge carrier \cite{Behniabook}.
The observed temperature dependence of $S/T$ 
($S$ and $T$ being the Seebeck coefficient and temperature, respectively)
has a maximum near 
characteristic temperatures in the magnetization
in both Co- and Mn-substituted samples,
indicating that both FM and AFM fluctuations are 
responsible for the enhancement of the Seebeck coefficient.
Moreover, 
in sharp contrast to the typical metallic behavior in which 
$S/T$ remains constant,
$S/T$ increases with cooling  in the parent compound,
implying
both FM and AFM fluctuations remaining in the itinerant electrons in Sr$_2$RuO$_4$. 

\section{II. Experimental details}

Single crystals of Sr$_2$Ru$_{1-x}M_x$O$_4$ ($M = $ Co, Mn) were
grown by a floating-zone method using an image furnace with a pair of halogen lamps and elliptical mirrors \cite{Mao1999, Ortmann2013}. 
We used high-purity SrCO$_3$(99.99\%+), RuO$_2$(99.9\%), MnO$_2$(99.99\%), and CoO(99.9\%) as starting materials. 
An excess 15\% amount of RuO$_2$ was weighed to compensate for evaporation of Ru during single crystal growth.
The concentration of Mn and Co in obtained crystals were determined by electron probe micro analyzer.
The measured Mn concentration was 2.9\%, well corresponding to the nominal concentration of 3\%.
In contrast, the measured Co concentrations were 1.8\% and 0.8\% for nominally 3\% and 1\% Co-substituted samples, respectively.
The relation between the nominal and measured concentrations is consistent with the trend reported in Ref. \onlinecite{Ortmann2013}.

\begin{figure}[t]
\begin{center}
\includegraphics[width=1\linewidth]{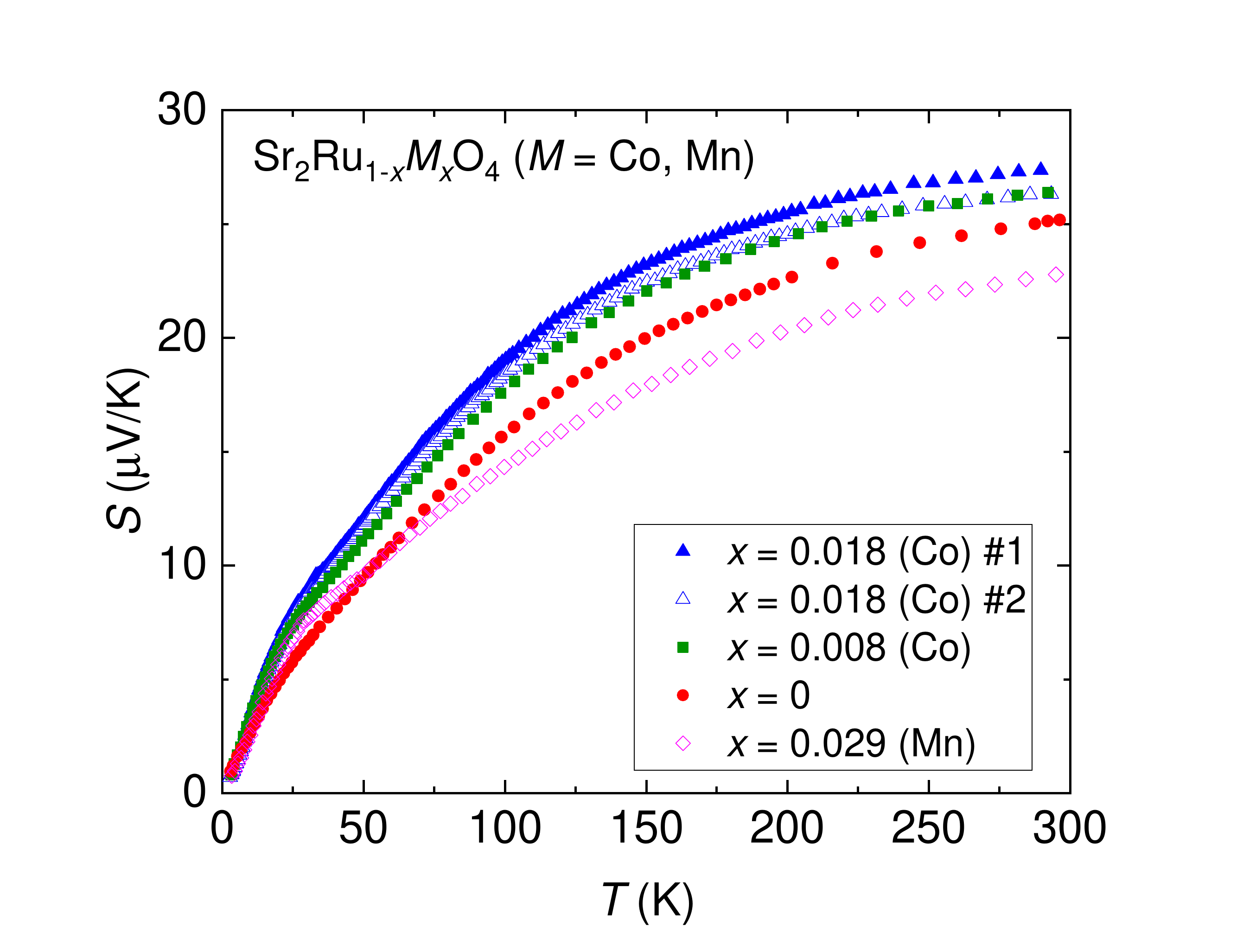}
\caption{Temperature dependence of the Seebeck coefficient $S$ of Sr$_2$Ru$_{1-x}M_x$O$_4$ ($M = $ Co, Mn) single crystals.}
\label{fig1}
\end{center}
\end{figure}

The typical dimension of the measured crystals was $\approx 3\times 0.2 \times 0.2$~mm$^3$.
The Seebeck coefficient was measured by a steady-state technique
using a manganin-constantan differential thermocouple in a closed-cycle refrigerator.
The thermoelectric voltage of the sample was measured with Keithley 2182A nanovoltmeter. 
The temperature gradient with a typical temperature gradient of 0.5 K/mm was applied along the $ab$-plane direction using a resistive heater.
The thermoelectric voltage from the wire leads was subtracted. 
Magnetization was measured using a superconducting quantum interference device magnetometer (Quantum Design, MPMS)
under field-cooled (FC) and zero-field-cooled (ZFC) processes with the external field of $\mu_0H=0.1$~T 
applied along the $c$ axis.

\section{III. results and discussions}

Figure 1 summarizes the temperature variations of the Seebeck coefficient $S$ in Sr$_2$Ru$_{1-x}M_x$O$_4$ ($M = $ Co, Mn) single crystals.
Overall behavior of $S$ in the parent compound Sr$_2$RuO$_4$ is consistent with earlier results \cite{Yoshino1996,Xu2008},
and is discussed as an intriguing example to study the internal degrees of freedom in correlated metals \cite{Mravlje2016}.
In the substituted compounds, 
$S$ increases (decreases) with Co (Mn) substitutions near room temperature
probably because of the electron (hole) doping effect
as suggested in Ref. \onlinecite{Ortmann2013}.

\begin{figure}[t]
\begin{center}
\includegraphics[width=1\linewidth]{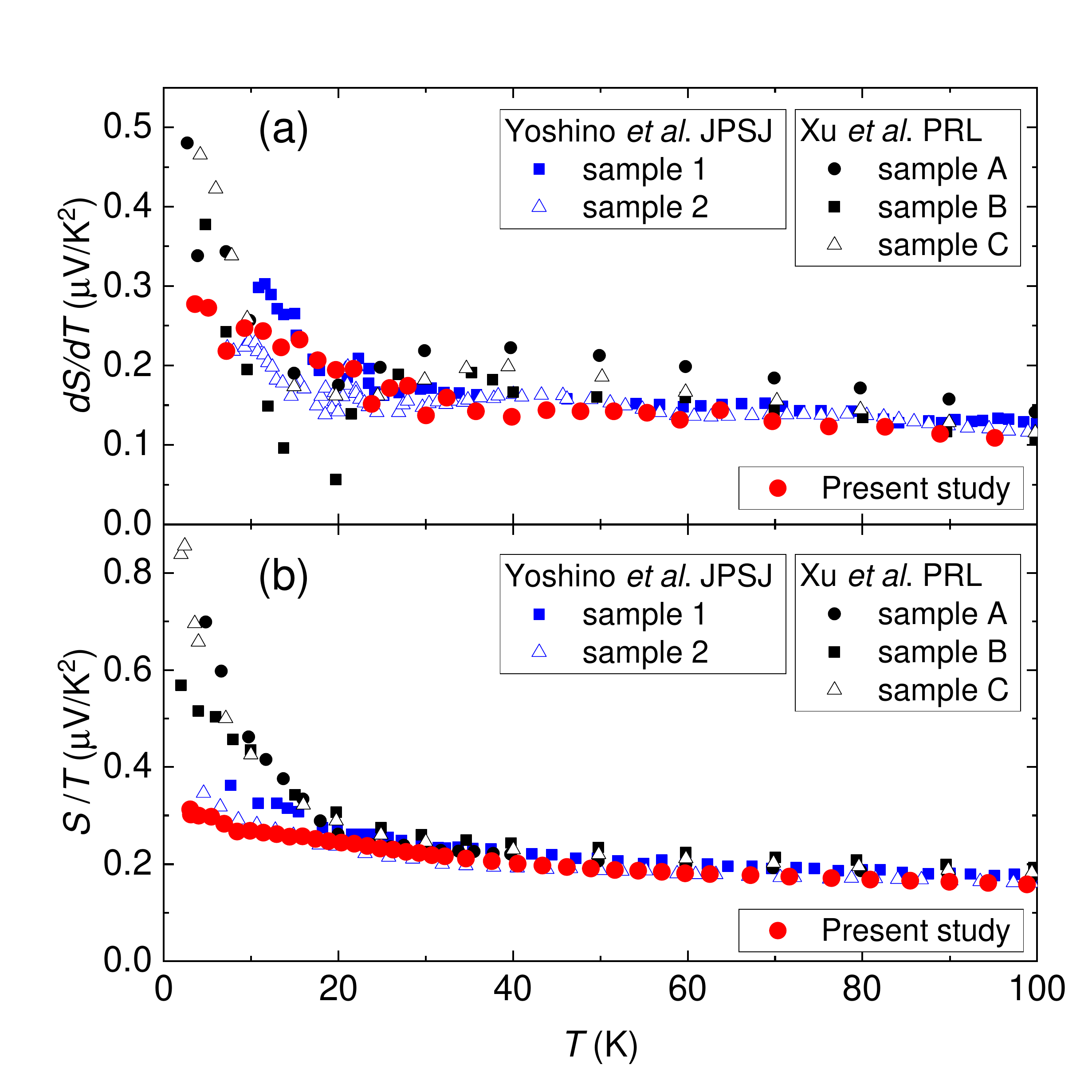}
\caption{
Comparison of the temperature dependence of (a) $dS/dT$ and (b) $S/T$ of Sr$_2$RuO$_4$ single crystals.
The red circles show the present data, and the others are data
extracted from Refs. \onlinecite{Yoshino1996,Xu2008}.}
\label{fig2}
\end{center}
\end{figure}

In Refs. \onlinecite{Yoshino1996,Xu2008}, 
the Seebeck coefficient in the parent compound was analyzed in the differential form $dS/dT$, and 
an anomaly was found near $20-25$~K. 
Xu \textit{et al.} have suggested that 
a band-dependent coherence is developed below the anomaly temperature
on the basis of the results of the Seebeck and the Nernst measurements \cite{Xu2008}.
Indeed, such a coherency seems to be vital in this system \cite{Mravlje2011}.
To see this anomaly at $20-25$~K, 
we compare the present data for Sr$_2$RuO$_4$ with the results extracted from Refs. \onlinecite{Yoshino1996,Xu2008} in Fig. 2(a).
Although there is a sample dependence in magnitude, 
all the data exhibit a kink around $20-25$~K,
in good agreement with the present result.
Also note that the effect from the impurity phase of SrRuO$_3$ is negligible,
because the Seebeck coefficient $S$ in a composite sample with the conductivity $\sigma$ is given as $\sigma S = \sum_k\alpha_k\sigma_kS_k$ in a parallel-circuit model,
where $\alpha_k,\sigma_k,S_k$ are the volume fraction, the conductivity, and the Seebeck coefficient for the material $k$ \cite{Okazaki2012}, and
the volume fraction of the impurity phase in the present crystal was negligibly small.

Here we discuss the Seebeck coefficient in the form of $S/T$ instead of $dS/dT$,
since the Seebeck coefficient in the free-electron model (oversimplified model to multiband Sr$_2$RuO$_4$) is expressed as $S\approx -\frac{k_{\rm{B}}^2T}{e}\frac{N(\mu)}{n}$,
where $k_{\rm{B}}$, $e$, $n$, $\mu$, and $N$ are the Boltzmann constant, elementary charge, carrier density, chemical potential, and the DOS, respectively \cite{Behnia2004}.
In this form, one can follow the temperature dependence of the DOS or carrier density,
as is widely analyzed in correlated electron systems,
similar to the case of the pseudogap state in transition-metal oxides \cite{Terasaki2001,Ikeda2016,Collignon2021} and 
heavy-fermion formation in rare-earth compounds \cite{Zlatic2003}.
Note that the differential $dS/dT$ includes the temperature derivatives of $N(\mu)$ and $n$ in complicated forms.
Figure 2(b) represents the temperature dependence of $S/T$ in Sr$_2$RuO$_4$ in the present crystals and 
the calculated $S/T$ from the extracted data from Refs. \onlinecite{Yoshino1996,Xu2008}.
Interestingly, all the data are temperature-dependent;
in sharp contrast to conventional metals, 
in which $S/T$ remains constant \cite{Behniabook},
$S/T$ significantly increases with decreasing temperature,
the origin of which will be discussed later.

\begin{figure}[t]
\begin{center}
\includegraphics[width=1\linewidth]{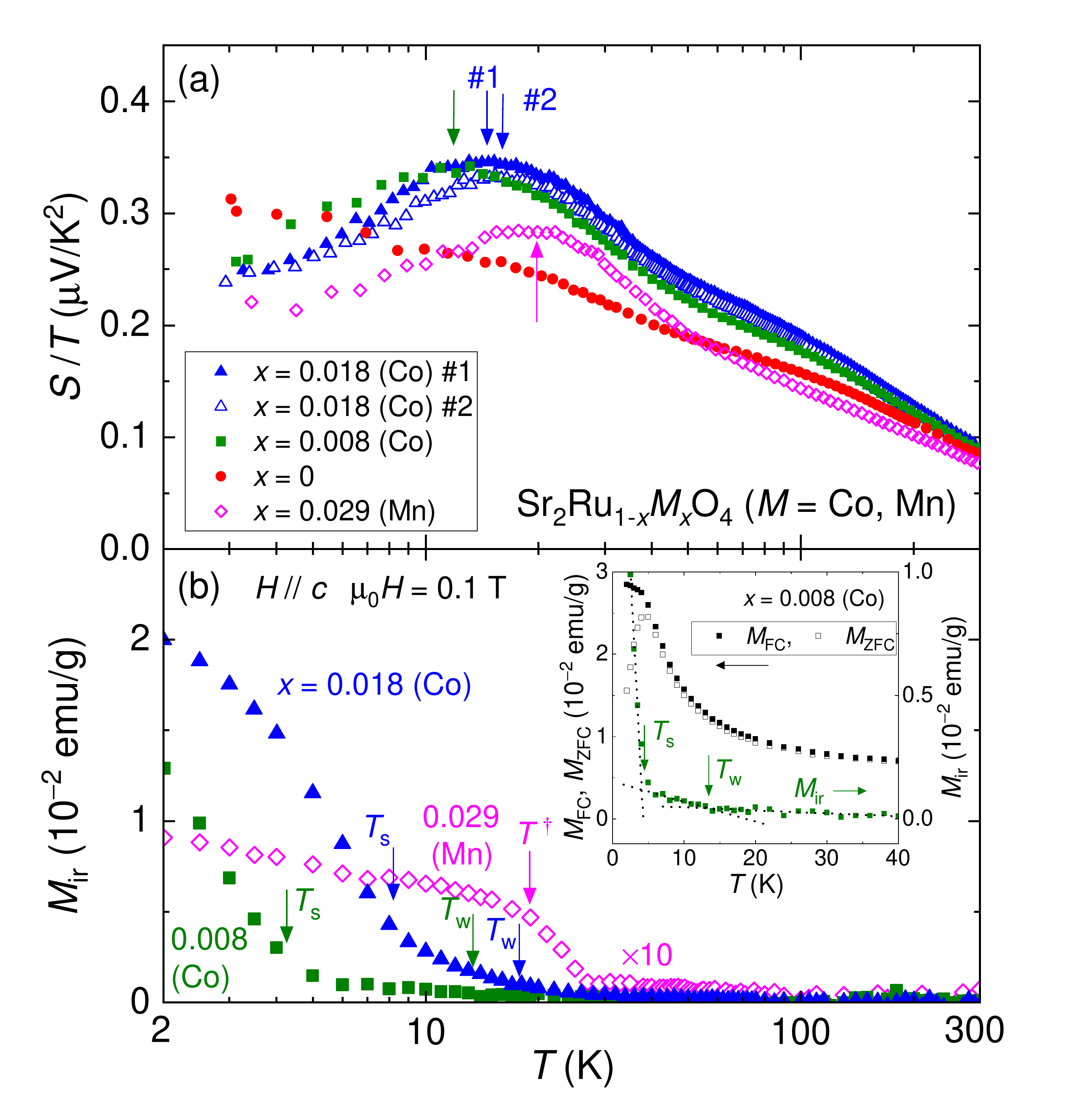}
\caption{
(a) Temperature dependence of $S/T$ of Sr$_2$Ru$_{1-x}M_x$O$_4$ ($M = $ Co, Mn) single crystals.
The arrows show the peak temperature $T_{\rm peak}$.
(b) Temperature dependence of
the irreversible magnetization $M_{\rm ir} \equiv M_{\rm FC}-M_{\rm ZFC}$.
The arrows represent the characteristic temperatures $T^\dagger$, $T_s$, and $T_w$. (See text for details.)
In the inset, $M_{\rm FC}$, $M_{\rm ZFC}$, (left axis) and $M_{\rm ir}$ (right axis) of the 0.8\% Co-substituted sample are shown 
together with the definitions of $T_{\rm s}$ and $T_{\rm w}$.
}
\label{fig3}
\end{center}
\end{figure}

We then focus on the temperature variation of $S/T$ in the substituted systems.
Figure 3(a) represents the temperature dependence of $S/T$ in 
Sr$_2$Ru$_{1-x}M_x$O$_4$ ($M = $ Co, Mn),
in which we find a drastic change due to substitutions.
At low temperatures, $S/T$ exhibits 
a prominent peak structure in the temperature dependence for both substituted systems.
The peak temperature $T_{\rm peak}$ defined as the temperature at which $S/T$ exhibits a maximum is shown by arrows in Fig. 3(a).
Note that the phonon-drag effect is unlikely to enhance the Seebeck coefficient in the substituted systems,
because the substitutions generally 
suppress the phonon mean free path \cite{Kurita2019}.

To shed light on the relation to the magnetism, 
we plot 
the irreversible magnetization $M_{\rm ir}$ defined as $M_{\rm ir}\equiv M_{\rm FC} - M_{\rm ZFC}$,
in Fig. 3(b).
($M_{\rm FC}$ and $M_{\rm ZFC}$ being the magnetization measured under FC and ZFC processes, respectively).
  While the onset of the irreversibility may correspond to a transition to the FM and AFM glassy states for
the Co- and Mn-substituted systems, respectively \cite{Ortmann2013},
a close look of the $M_{\rm ir}$ data reveals that the onset of the irreversibility is accompanied by a double step.
The inset of Fig. 3(b) shows the temperature variations of $M_{\rm FC}$, $M_{\rm ZFC}$, and $M_{\rm ir}$ for $x=0.008$ (Co) at low temperatures.
The $M_{\rm ZFC}$ data more steeply decrease than that in Ref. \onlinecite{Ortmann2013}, 
but this may be due to the low applied field in our measurement.
From the plots of $M_{\rm ir}$ in the inset, 
one can see that a weak irreversibility sets in at $T_{\rm w}\approx14$~K
and subsequently the irreversibility is strongly enhanced below $T_{\rm s}\approx 4$~K.
The weak and strong irreversibility temperatures $T_{\rm w}$ and $T_{\rm s}$ are defined as the points of intersection of 
the two linear lines as drawn in the dotted line in the inset.
Importantly,
such a coexistence of weak and strong irreversibility is often observed in ferromagnetic spin-glass systems \cite{Pappa1984,DeFotis1998,Dahr2006,Lago2012},
and theoretically interpreted in terms of a multi component vector spin model given by Gabay and Toulouse \cite{Gabay1981},
in which the transverse spin components are first frozen at $T_{\rm w}$ on cooling
and then followed by the freezing of the longitudinal spin components at $T_{\rm s}$,
although it is still under debate whether it represents a true thermodynamic phase transition.

On the other hand, the temperature dependence of $M_{\rm ir}$ in the Mn-substituted sample 
is different from that in the Co-substituted samples:
On cooling, the irreversibility sets in at $T\approx27$~K,
which is almost identical to the onset temperature $T_\mathrm{ir}$ to the static order \cite{Ortmann2013},
 and then shows weak temperature dependence below $T^\dagger \approx18$~K
at which $M_{\rm ir}$ displays a kink structure as shown in Fig. 3(b). 
In the Mn-substituted sample,
$T_\mathrm{peak}$ in $S/T$ is close to this anomaly temperature $T^\dagger$,
although the origin of $T^\dagger$ is unclear at present.
Note that the remanent magnetization is also distinct among the Co- and Mn-substituted systems:
While the temperature dependence of remanent magnetization is monotonic in the Co-substituted samples,
it exhibits an anomalous peak structure below the onset temperature $T_\mathrm{ir}$ for the Mn-substituted systems  \cite{Ortmann2013},
implying the existence of a characteristic temperature below $T_\mathrm{ir}$.
The nature of the magnetic glassy state is an issue for the future study using
the microscopic NMR and neutron measurements.

\begin{figure}[t]
\begin{center}
\includegraphics[width=0.9\linewidth]{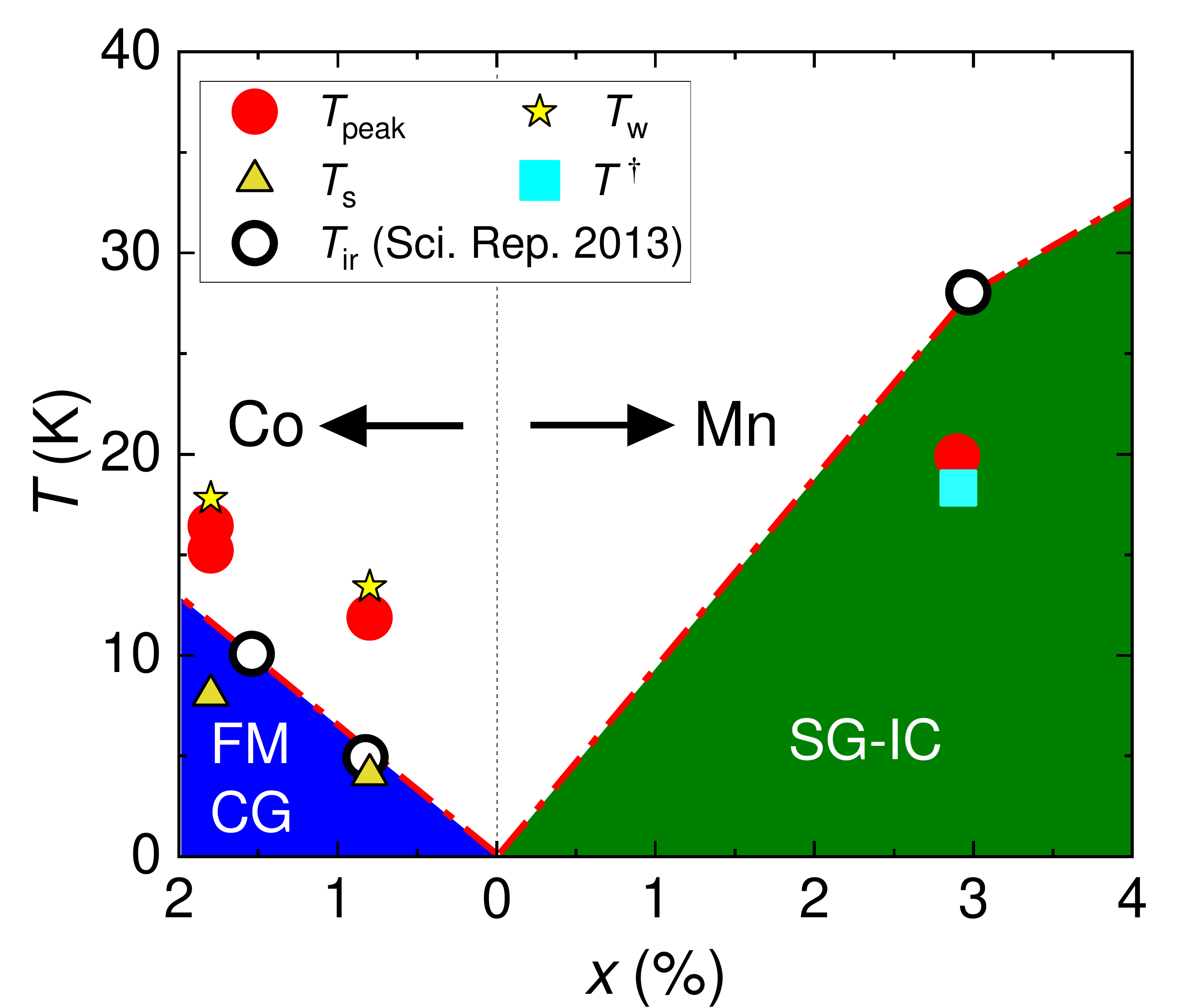}
\caption{
Phase diagram of Sr$_2$Ru$_{1-x}M_x$O$_4$ ($M = $ Co, Mn).
The closed circles shows the peak temperature $T_{\rm peak}$ at which $S/T$ exhibits maximum.
The weak and strong irreversibility temperatures $T_{\rm w}$ (stars) and $T_{\rm s}$ (triangles) and 
the anomaly temperature $T^\dagger$ (square) are determined from the irreversibility magnetization $M_{\rm ir}$. 
The onset temperature $T_{\rm ir}$ (open circles)
and the regions of FM cluster glass (CG) and spin glass state with short-range incommensurate AFM order (SG-IC) are extracted from Ref. \onlinecite{Ortmann2013}.
}
\label{fig4}
\end{center}
\end{figure}

In Fig.~\ref{fig4}, we plot the peak temperature $T_{\rm peak}$ in $S/T$ 
together with $T_\mathrm{s}$, $T_\mathrm{w}$, and $T^\dagger$ on the phase diagram reported in 
Ref.~\onlinecite{Ortmann2013}.
The peak temperature $T_\mathrm{peak}$ in $S/T$ well coincides with the weak irreversibility temperature $T_\mathrm{w}$
and the anomaly temperature $T^\dagger$ 
for the Co- and Mn-substituted samples, respectively,
indicating an intimate relationship among the Seebeck coefficient and the magnetic fluctuations.
Here, 
since the Co and Mn substitutions induce 
the FM cluster glass and the spin glass states with short-range AFM order, respectively \cite{Ortmann2013},
the present results 
indicate that 
both FM and AFM fluctuations are substantial for the enhancement of the Seebeck coefficient at $T_{\rm peak}$.
Indeed, it bears a striking resemblance to that in the itinerant magnets such as
the perovskite ruthenate CaRu$_{0.8}$Sc$_{0.2}$O$_3$ \cite{Yamamoto2017}
and 
the doped Heusler alloy Fe$_2$VAl \cite{Tsujii2019};
in these compounds, 
the Seebeck coefficient is enhanced near the ferromagnetic transition temperature 
through a sort of the magnon-drag effect,
and 
is reduced by applying magnetic field owing to the field-induced suppression of the magnetic fluctuation.
In the antiferro-quadrupole (AFQ) PrIr$_2$Zn$_{20}$, moreover, $S/T$ exhibits a peak structure near the AFQ transition temperature in high magnetic field where the quadrupolar fluctuation could be enhanced\cite{Ikeura2013}. 

Now let us recall the $S/T$ behavior in Sr$_2$RuO$_4$ (Fig. 2),
which increases with cooling down to the lowest temperature of the present measurement ($\sim3$~K).
While Seebeck coefficients are not strictly linear in temperature even in simple metals \cite{Macdonald},
the non-linearity in Sr$_2$RuO$_4$ is unusual. 
As mentioned above, we show the suggestive evidence that both FM and AFM fluctuations 
drive the enhanced Seebeck coefficient in the vicinity of the parent compound.
It is therefore reasonable to consider that, even in the parent compound Sr$_2$RuO$_4$,
$S/T$ is enhanced similarly down to zero temperature,
at which both FM and AFM glassy states tend to terminate for Sr$_2$RuO$_4$\cite{Ortmann2013}.
Thus this is a kind of non-Fermi-liquid (NFL) behavior close to the quantum critical point \cite{Paul2001},
which has also been observed in various correlated matters such as 
heavy fermions \cite{Izawa2007,Hartmann2010,Malone2012,Pfau2012,Shimizu2015} and
oxides \cite{Limelette2010}.
Quite intriguingly, in contrast to the aforementioned systems,
both FM and AFM fluctuations seem to be substantial for the itinerant electrons in Sr$_2$RuO$_4$
as
seen in the pronounced peak of $S/T$ observed in both Co- and Mn-substituted compounds.

One may, however, pose a simple question regarding the specific heat.
In Sr$_2$RuO$_4$,
the electronic specific heat shows conventional Fermi-liquid (FL) behavior \cite{Maeno1997},
in distinction to the NFL behavior observed in the Seebeck coefficient;
since 
the Seebeck coefficient is also given as the specific heat per carrier \cite{Behnia2004},
both quantities are expected to show similar anomalies \cite{Kim2010}.
On the other hand, 
an AFM quantum criticality may affect the ratio of these quantities in zero-temperature limit \cite{Miyake2005}.
It is interesting to note that this effect of the AFM quantum criticality may be enhanced by 
the multiband nature; 
in Sr$_2$RuO$_4$,  
the Fermi surfaces are composed of three cylindrical sheets: 
hole-like $\alpha$ and electron-like $\beta$ and $\gamma$ sheets 
with the order of magnitudes of the effective mass of $m_{\alpha}<m_{\beta}<m_{\gamma}$ 
\cite{Oguchi1995,Singh1995,Mackenzie1996,Hill2000}.
Since the lighter band contributes to the transport more significantly in general \cite{Kasahara2007,Yano2008}, 
the lighter $\alpha$ and $\beta$ bands
are essential here, and importantly,
these$\alpha$ and $\beta$ sheets
possess the AFM instability due to nesting \cite{Sidis1999,Braden2002,Steffens2019,Jenni2021}.
This band-dependent magnetic fluctuation may give origin to
the anomalous low-temperature increase in $S/T$.
On the other hand, 
it is unclear at present how the thermoelectric transport is affected by
the FM fluctuation, the importance of which is clearly demonstrated in the Co-substituted systems;
the existence of FM fluctuation is indicated by NMR \cite{Imai1998}, 
while the neutron experiments have revealed that it is very weak \cite{Sidis1999,Braden2002,Steffens2019,Jenni2021}.

It is also known that the electrical resistivity of Sr$_2$RuO$_4$ is well described within the FL scheme \cite{Maeno1997},
in which the resistivity is given as $\rho(T) = \rho_0+AT^\nu$ with the exponent $\nu=2$.
On the other hand, the determination of $\nu$ is delicate \cite{Kikugawa2002}, 
and as seen in other correlated materials, 
$\nu$ apparently depends on the residual resistivity even in high-purity level \cite{Tateiwa2012}.
It may be useful to examine the exponent in high-purity crystals \cite{Bobowski2019}.
We also mention that the thermal conductivity in the normal state is interesting 
in the sense that it also mirrors the entropy flow, 
while the earlier studies are mainly devoted to the superconducting state \cite{Tanatar2001,Izawa2001,Suzuki2002}.
Also, the Seebeck coefficient in the system with the Fermi level at the vHs, which is not achieved with the present Co substitution,
is worth exploring
because the topological change in the Fermi surface at vHs 
may lead to the significant enhancement of the Seebeck coefficient \cite{Varlamov1989,Ito2019}.
This effect of the vHs could be investigated in electron-doped Sr$_{2-x}$La$_x$RuO$_4$.

\section{IV. Conclusion}

  To summarize, we performed Seebeck coefficient measurement in Sr$_2$Ru$_{1-x}M_x$O$_4$ ($M = $ Co, Mn). 
  Although the $S/T$ of the parent compound Sr$_2$RuO$_4$ increases with cooling down to the lowest temperature $\sim3$~K similar to previous reports,
  for Co- and Mn-substituted systems, $S/T$ is enhanced near
  $T_{\rm w}$ and $T^\dagger$ in the irreversible magnetization. 
  The emergence of the peak structure in $S/T$ can be related to glassy FM and AFM fluctuations in the Co- and Mn-substituted system, respectively,
  and therefore the increase of $S/T$ in Sr$_2$RuO$_4$ persisting at least down to $\sim3$~K suggests that both FM and AFM 
  fluctuations seem to be substantial in Sr$_2$RuO$_4$.

\begin{acknowledgments}
The authors acknowledge Y. Maeno and K. Ishida for fruitful discussions.  
This work was supported by JSPS KAKENHI Grants No. JP17H06136 and No. JP18K13504.
\end{acknowledgments}


\end{document}